\documentclass[aps,prb,twocolumn,superscriptaddress,showpacs]{revtex4}
\usepackage{epsfig}
\usepackage{graphicx}
\usepackage{dcolumn}
\usepackage{bm}

\begin{document}

\title{Effect of electron correlations on the electronic structure and phase stability of FeSe upon lattice expansion}

\author{S. L. Skornyakov}

\affiliation{Institute of Metal Physics, Sofia Kovalevskaya Street 18, 620219 Yekaterinburg GSP-170, Russia}
\affiliation{Ural Federal University, 620002 Yekaterinburg, Russia}

\author{V. I. Anisimov}

\affiliation{Institute of Metal Physics, Sofia Kovalevskaya Street 18, 620219 Yekaterinburg GSP-170, Russia}
\affiliation{Ural Federal University, 620002 Yekaterinburg, Russia}

\author{D. Vollhardt}

\affiliation{Theoretical Physics III, Center for Electronic Correlations and
Magnetism, Institute of Physics, University of Augsburg, 86135 Augsburg, Germany}

\author{I. Leonov}
\affiliation{Theoretical Physics III, Center for Electronic Correlations and
Magnetism, Institute of Physics, University of Augsburg, 86135 Augsburg, Germany}

\affiliation{Materials Modeling and Development Laboratory, National University of Science and Technology 'MISIS', 119049 Moscow, Russia}

\date{\today}

\begin{abstract}
We present results of a detailed theoretical study of the electronic, magnetic, and structural
properties of the chalcogenide parent system FeSe using a fully charge self-consistent implementation
of the density functional theory plus dynamical mean-field theory (DFT+DMFT) method. In particular, we predict a
remarkable change of the electronic structure of FeSe which is accompanied by a complete reconstruction
of the Fermi surface topology (Lifshitz transition) upon a moderate expansion of the lattice volume.
The phase transition results in a change of the in-plane magnetic nesting wave vector from $(\pi,\pi)$
to $(\pi,0)$ and is associated with a transition from itinerant to orbital-selective localized magnetic
moments. We attribute this behavior to a correlation-induced shift of the van Hove singularity of the
Fe $t_{2}$ bands at the M-point across the Fermi level.
Our results reveal a strong orbital-selective renormalization of the effective mass $m^*/m$ of the Fe
$3d$ electrons upon expansion. The largest effect occurs in the Fe $xy$ orbital, which gives rise to a
non-Fermi-liquid-like behavior above the transition.
The behavior of the momentum-resolved magnetic susceptibility $\chi({\bf q})$ demonstrates that magnetic
correlations are also characterized by a pronounced orbital selectivity, suggesting a spin-fluctuation origin of the nematic phase
of paramagnetic FeSe.
We conjecture  that the anomalous behavior of FeSe upon expansion is associated with the
proximity of the Fe $t_{2}$ van Hove singularity to the Fermi level and the sensitive dependence of its
position on external conditions.

\end{abstract}

 \maketitle

%%%%%%%%%%%%%%%%%%%%%%%%%%%%%%%%%%%%%%%%%%%%%%%%%%%%%%%%%%%%%%%%%%%%%%%%%%%%

\section{Introduction}

During the last decade the electronic, magnetic, and structural 
properties of the iron-based high-temperature superconducting 
pnictides and chalcogenides have been the subject of intensive 
research\cite{pnictide_discovery,review_superconductors}. These 
novel superconducting materials show certain similarities with 
the high-$T_c$ cuprate superconductors. Indeed, the iron-based 
superconductors (FeSC) adopt a quasi 2D crystal structure where 
the iron atoms form a square lattice. The latter are separated 
by non-conducting layers containing, for example, alkali, alkaline 
earth, or rare earth elements, oxygen and/or fluorine. Moreover, 
the superconducting phase of these novel compounds often appears 
in the vicinity of a magnetic phase transition and/or structural 
instability. In particular, superconductivity in FeSCs often 
occurs as a result of the suppression  of long-range, single-stripe 
antiferromagnetic (AF) order with a wave vector $Q_m = (\pi, \pi)$, 
due to electron/hole doping or pressure. This behavior has been 
regarded as evidence for the importance of spin fluctuations in 
the pairing of electrons in FeSCs.

The newly discovered Fe$_{1+y}$Se is structurally the simplest 
among the FeSCs \cite{FeSe_structure}. At ambient pressure it has 
been found to become superconducting below $T_{c} \sim$ 8~K close 
to its stoichiometric composition \cite{Superconductivity_FeSe}.
FeSe has the same layered structure as the pnictides, containing 
layers of edge-sharing FeSe$_4$ tetrahedra, but without separating 
(non-conducting) layers \cite{FeSe_structure}. Therefore FeSe is 
viewed as the parent compound of Fe-based superconductors which 
represents a minimal model material for understanding the mechanism 
of superconductivity of FeSCs. Moreover, FeSe itself exhibits 
remarkable physical properties. Its critical temperature $T_c~\sim$~8~K 
at normal pressure increases to $\sim$~14~K upon isovalent substitution 
of Se with Te (corresponding to a negative chemical pressure, i.e., 
lattice expansion \cite{FeSe_Te_doping}), to $\sim$~37~K under 
compression \cite{FeSe_hydrostatic}, to $\sim$~40~K by means of 
intercalation \cite{FeSe_intercalation}, and all the way up to 
$\sim$~65-109~K in the case of a monolayer \cite{FeSe_monolayer}.
In addition, FeSe has been found to exhibit a transition to a nematic 
phase below $\sim$~90~K  in which the crystal (C$_4$) rotation 
symmetry is spontaneously broken \cite{nematicity}. Due to these 
intriguing properties FeSe has attracted much recent attention 
from both theory and experiment.

Angle-resolved photoemission spectroscopy (ARPES) \cite{FeSe_photoemission_1,FeSe_photoemission_2,FeSe_photoemission_3}
and band structure calculations \cite{DFT_FeSe} reveal that the 
electronic structure of FeSe resembles that of the pnictides. It 
has a quasi 2D Fermi surface with three concentric hole pockets 
at the Brillouin zone $\Gamma$-point and two intersecting elliptical 
electron pockets centered at the M-point. The Fermi surface topology 
is characterized by an in-plane nesting wave vector $(\pi,\pi)$, 
consistent with $s^{\pm}$ pairing symmetry \cite{s_pm_pairing}. 
Moreover, experimental studies of the spin excitation spectra of 
both pnictides and chalcogenides show an enhancement of short-range
AFM spin fluctuations at vector $(\pi,\pi)$ near the $T_c$ \cite{FeSe_pi_pi_fluctuations}.
These results suggest a common origin of superconductivity in 
pnictides and chalcogenides, for example due to spin fluctuations 
associated with the suppression of long-range magnetic order.

Unlike the majority of the FeSCs, FeSe is not magnetically ordered 
at ambient pressure and composition \cite{FeSe_phase_diag, FeSe_magnetism}. 
Its isoelectronic counterpart FeTe, the end member of the Fe(Se,Te) 
series, is antiferromagnetic below the N\'eel temperature of 70~K. 
However, in contrast to the magnetic phases of the Fe-pnictides, 
FeTe exhibits double-stripe AF order with a $(\pi,0)$ propagation 
vector \cite{FeSe_T_cT}. Upon compression, FeTe exhibits a transition 
to a collapsed-tetragonal phase which is accompanied by a collapse 
of magnetic moments \cite{FeSe_T_cT}. All this suggests a reconstruction 
of the electronic structure of Fe(Se,Te) upon change of the Se content 
or compression.

Photoemission and ARPES measurements of the electronic properties 
of Fe(Se,Te) reveal a significant narrowing of the Fe $3d$ bandwidth
as compared to band structure calculations \cite{FeSe_photoemission_1}.
This corresponds to a strong orbital-dependent enhancement of the 
quasiparticle mass in the range $\sim$ 3-20 compared with the values 
obtained by electronic band structure techniques \cite{FeSe_photoemission_2,FeSe_photoemission_3}.
Moreover, these experiments exhibit a damping of the coherent 
quasiparticles in Te-rich Fe(Se,Te), indicating a crossover from a 
coherent to incoherent behavior of the electronic structure.
In addition, with increasing Te content, ARPES data for Fe(Se,Te) 
show a suppression of the spectral weight intensity associated with 
a Fermi surface pocket at the Brillouin zone M-point \cite{FeSe_photoemission_3}. 
This behavior is accompanied by an enhancement of spectral weight 
at the X-point, implying a possible doping-induced reconstruction 
of the electronic structure. Overall these experimental results 
point towards the importance of strong orbital-selective correlations.

State-of-the-art methods for the calculation of the electronic 
properties of strongly correlated systems, such as the density 
functional theory plus dynamical mean-field theory (DFT+DMFT) 
approach \cite{dmft,dftdmft} provide a good qualitative and even 
quantitative description of the band structure of FeSCs \cite{U_in_superconductors}.
Applications of DFT+DMFT to FeSe yield a band mass enhancement 
in the range 2-5 and, in contrast to the pnictides, reveal the 
presence of a lower Hubbard band in the spectral function of 
FeSe \cite{FeSe_Aichhorn_2010,WB16}. This clearly demonstrates 
the importance of correlation effects for the electronic properties 
of FeSe. Moreover, our recent DFT+DMFT calculations of the electronic 
properties and phase stability of FeSe predict that FeSe undergoes 
a phase transformation from a collapsed tetragonal to tetragonal 
phase upon expansion of the lattice \cite{FeSe_Leonov_2015}.
The transformation is found to be accompanied by a complete 
reconstruction of the topology of the Fermi surface (Lifshitz transition), 
a sharp increase of the local moments, and a change of magnetic 
correlations due to a transition of the in-plane magnetic wave 
vector from $(\pi,\pi)$ to $(\pi,0)$. This behavior was attributed 
to a correlation-induced shift of the van Hove singularity 
associated with the Fe $xy$ and $xz/yz$ orbitals at the Brillouin
zone M-point across the Fermi level \cite{FeSe_Leonov_2015}.

The present study extends our previous investigation of FeSe 
\cite{FeSe_Leonov_2015}. In particular, we now perform fully 
charge self-consistent DFT+DMFT calculations to determine the 
electronic properties and phase stability of paramagnetic FeSe. 
To this end, we take the crystal structure data for the 
paramagnetic tetragonal phase of FeSe from experiment \cite{FeSe_structure} 
and calculate the total energy as a function of volume. Our 
results reveal a substantial change of the total energy upon 
inclusion of the effects of charge redistribution caused by 
correlation effects. This proves the general importance of 
electronic correlations on the charge density and, hence, on 
the orbital occupancies. While this influence turns out to 
be negligible for the equilibrium volume, it becomes significant 
at higher volumes. At the same time the actual results for 
the electronic structure and phase stability show no qualitative 
difference compared to those calculated without charge 
self-consistency \cite{FeSe_Leonov_2015}. Namely, the fully 
charge self-consistent calculations still find a structural 
phase transition upon expansion of the lattice, which is 
associated with a reconstruction of the topology of the Fermi 
surface (Lifshitz transition) and is accompanied by a sharp 
increase of the local moments. Indeed, our analysis of the 
Fermi surface topology and results for the spin susceptibility 
$\chi({\bf q})$ support the previously suggested reconstruction 
of magnetic correlations from the in-plane magnetic wave vector 
$(\pi,\pi)$ to $(\pi,0)$, indicating a competition between 
these two magnetic instabilities \cite{NatComm.7.12182}. 
Moreover, we find that the individual orbitals contribute 
very differently to $\chi({\bf q})$, a fact which may play 
a crucial role in explaining the observed nematicity in 
Fe(Se,Te) compounds \cite{nematicity}. Our calculations 
reveal a pronounced orbital-selective enhancement of 
electronic correlation upon expansion of the lattice. 
In particular, we observe a crossover from a Fermi-liquid 
with a weak self-energy-induced damping at the Fermi energy, 
to a non-Fermi-liquid like behavior where the self-energy 
almost diverges. The crossover is found to be associated 
with a transformation from an itinerant to localized magnetic 
moment behavior. Our results clearly demonstrate the crucial 
importance of orbital-selective correlations for a realistic 
description of the electronic and lattice properties of FeSe.

%%%%%%%%%%%%%%%%%%%%%%%%%%%%%%%%%%%%%%%%%%%%%%%%%%%%%%%%%%%%%%%%%%%

\section{Method}
In this paper, we employ a state-of-the-art DFT+DMFT computational 
scheme \cite{dmft,dftdmft}, which is fully self-consistent in 
the charge density, to determine the electronic properties and 
phase stability of paramagnetic tetragonal FeSe. It is implemented 
\cite{LB08} within the non-spin polarized generalized gradient 
approximation (GGA) in DFT using plane-wave pseudopotentials 
\cite{pseudopotential}. The approach combines a construction of 
the low-energy Hamiltonian for the partially filled Fe $3d$ and 
Se $4p$ orbitals in the basis of Wannier functions \cite{WannierH} 
with the solution of the DMFT impurity problem using the 
continuous-time hybridization-expansion (segment) quantum Monte 
Carlo method \cite{ctqmc}. The effects caused by the correlation-induced 
charge redistribution are taken into account by solving the DFT+DMFT 
equations self-consistently in the charge density.

To investigate the structural stability, we use the atomic positions 
and the lattice parameter $c/a$ of paramagnetic tetragonal FeSe 
taken from experiment \cite{FeSe_structure}. To this end, we adopt 
the crystal structure data (space group $P4/nmm$, the lattice parameter 
ratio $c/a$=1.458, and the $z$-position of Se $z$=0.266) and calculate 
the total energy as a function of volume. In these calculations, 
we consider a uniform expansion or contraction of the lattice 
volume, i.e., only the lattice parameter $a$ is varied, while the 
$c/a$ ratio is fixed. We use the average Coulomb interaction 
$U$~=~3.5~eV and Hund's exchange $J$~=~0.85~eV for the Fe $3d$ shell, 
which are typical for the pnictides and chalcogenides according
to different estimations \cite{U_in_superconductors}. The Coulomb 
interaction is treated in the density-density approximation. 
The spin-orbit coupling is neglected in these calculations. 
We employ the fully localized double-counting correction, evaluated 
from the self-consistently determined local occupancies, to account 
for the electronic interactions already described by DFT. The spectral 
functions and angle resolved spectra are evaluated from analytic
continuation of the self-energy using Pad\' e approximants.

We analyze possible magnetic instabilities of FeSe by calculating 
the static momentum-dependent susceptibility $\chi({\bf q})$ within 
the particle-hole bubble approximation:
\begin{equation}
\chi({\bf q})=-k_{\mathrm{B}}\mathrm{T}\hspace{1mm}\mathrm{Tr}
\sum_{{\bf q},i\omega_{n}}\hat G({\bf k},i\omega_{n})\hat G({\bf k}+{\bf q},i\omega_{n}).
\label{eq_chiofq}
\end{equation}
Here $\mathrm{T}$ is the temperature, $\omega_{n}=(2n+1)\pi k_{\mathrm{B}}\mathrm{T}$ 
is the Matsubara frequency, $\hat G({\bf k},i\omega_{n})$ is the 
interacting lattice Green's function
\begin{equation}
\hat G({\bf k},i\omega_{n})=[(i\omega_{n}+\mu)\hat I - \hat H({\bf k})-\hat \Sigma(i\omega_{n})]^{-1},
\end{equation}
where $\mu$ is the chemical potential, $\hat H({\bf k})$ is the 
effective low-energy Hamiltonian in Wannier basis, and $\hat\Sigma(i\omega_{n})$ 
is the self-energy which includes an energy shift due to the 
double-counting correction term.

%%%%%%%%%%%%%%%%%%%%%%%%%%%%%%%%%%%%%%%%%%%%%%%%%%%%%%%%%%%%%%%%%%%%%%%%%%%%

\section{Results}

\subsection{Phase stability and local magnetic moments}

\begin{figure}[t]
\centering 
\includegraphics[width=0.4\textwidth,clip=true]{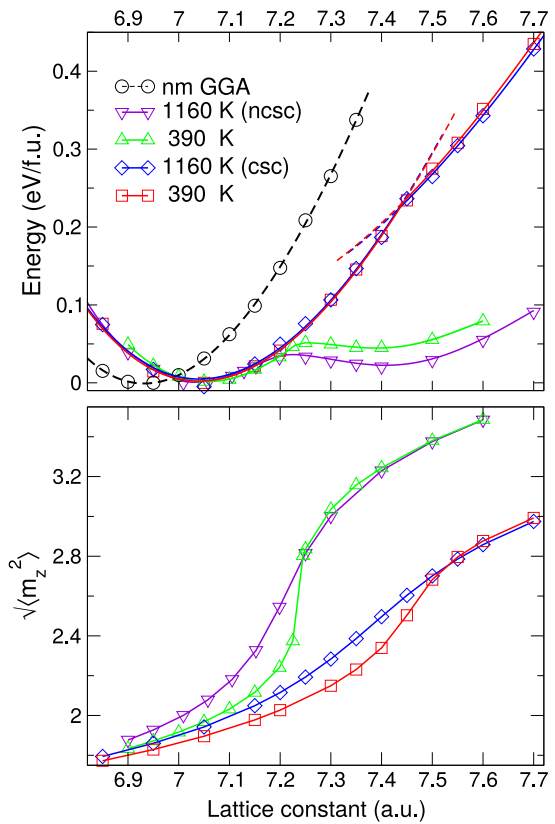}
\caption{(Color online)
Total energy (upper panel) and instantaneous local magnetic moments
$\sqrt{\langle m_z^2 \rangle}$ (lower panel) of paramagnetic FeSe as
a function of lattice constant calculated by DFT+DMFT at a temperature
$T=390$ K with (csc) and without (ncsc) charge self-consistency.
The total energy curve obtained with nonmagnetic GGA (nm GGA) is shown
in the upper panel for comparison.
}
\label{Fig_1}
\end{figure}

As a starting point, we compute the electronic structure and phase 
stability of paramagnetic FeSe. To this end, we evaluate the total 
energy of FeSe as a function of lattice volume by employing a fully 
charge self-consistent (csc) DFT+DMFT scheme \cite{V2O3_Leonov_2015,FOR1346_P2_review_2017}
and compare the result with that obtained from non-charge self-consistent
(ncsc) DFT+DMFT calculations \cite{FeSe_Leonov_2015} (Fig. \ref{Fig_1}).
The calculated equilibrium lattice constant $a=7.05$ a.u. at a temperature 
$T=390$~K is in good quantitative agreement with the experimental 
data \cite{FeSe_structure}, and to a good accuracy coincides with 
that obtained within ncsc DFT+DMFT \cite{FeSe_Leonov_2015}. We note 
that within the nonmagnetic generalized gradient approximation (GGA) 
the equilibrium lattice constant is substantially underestimated \cite{FeSe_Leonov_2015}.
We also observe a substantial change of the total energy when the 
correlation-induced charge redistribution is taken into account.
This clearly demonstrates the importance of the feedback of electronic 
correlations to the charge density. However, we find that this change 
is not very important for the actual value of the equilibrium volume.
It only becomes notable at larger volumes, where it results in a shift 
of a lattice anomaly from 7.25 a.u. in the ncsc calculation to 7.45 in 
the csc calculations. We also estimate the bulk modulus $K$ for the 
equilibrium phase by fitting the obtained energy-volume dependence 
using the third-order Birch-Murnaghan equation of state \cite{birch}. 
The computed value $K=79$ GPa and its pressure derivative $K' \equiv dK/dP=4.3$
at $T=390$ K are close to those obtained by the ncsc calculations \cite{FeSe_Leonov_2015}.
The computed instantaneous local magnetic moment $\sqrt{\langle m_z^2 \rangle}$ 
is about 1.9~$\mu_{B}$, corresponding to a fluctuating local magnetic 
moment of $\sim$ 0.7~$\mu_B$ \cite{note}. Clearly, it is the inclusion 
of the local Coulomb interaction that provides an overall improved 
description of the properties of FeSe compared to the DFT results.

\begin{figure}[t]
\centering
\includegraphics[width=0.4\textwidth,clip=true]{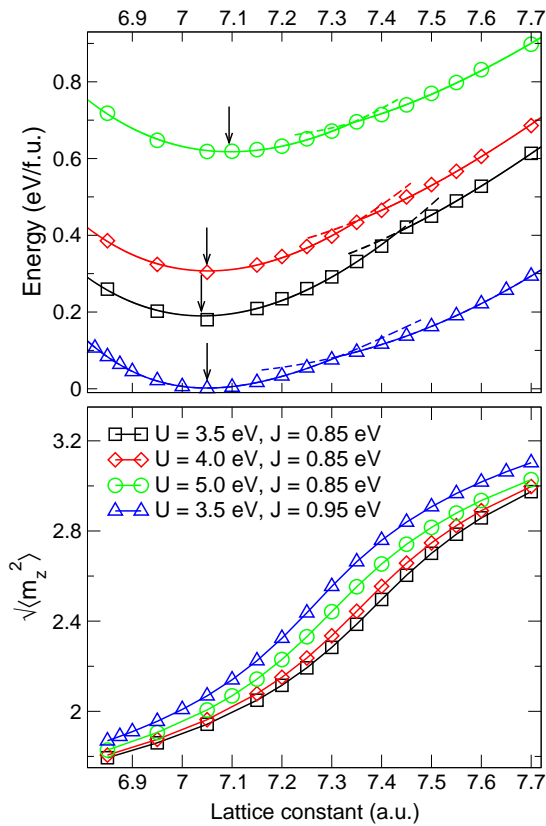}
\caption{(Color online)
Total energy (upper panel) and instantaneous local magnetic moments 
$\sqrt{\langle m_z^2 \rangle}$ (lower panel) of paramagnetic FeSe 
calculated for different interaction parameters $U$ and $J$ at $T=1160$~K 
using the charge self-consistent DFT+DMFT method. The arrows in the 
upper panel indicate the position of the energy minima.}
\label{Fig_2}
\end{figure}

Both in the ncsc and csc DFT+DMFT calculations the local magnetic 
moment is found to increase upon expansion of the lattice volume 
(Fig.~\ref{Fig_1}). We observe that charge self-consistency leads 
to a smoother evolution of the local moment and to a reduction of 
its absolute value in the whole range of lattice parameters.
Moreover, ncsc and csc calculations both predict an iso-structural 
phase transition which is associated with a substantial increase 
of the local magnetic moment $\sqrt{\langle m_z^2 \rangle}$ upon 
expansion of the lattice. In view of the experimental findings for 
the volume and local magnetic moment of FeTe upon compression \cite{FeSe_T_cT}, 
we interpret this behavior of FeSe as a transition from a 
collapsed-tetragonal (equilibrium volume) to tetragonal (expanded 
volume) phase which occurs upon expansion of the lattice. The 
expansion corresponds to a negative pressure of above $\sim$~-7.6~GPa. 
The expanded-volume phase has a significantly smaller bulk modulus 
of about 49 GPa. For $a=7.6$ a.u. the calculated local magnetic 
moment is $\sim 2.9~\mu_B$ (the fluctuating local moment is $\sim$2.6 
$\mu_B$). Our results show that the transition is accompanied by 
an increase of the lattice constant from $a=7.35$ a.u. to $a=7.6$~a.u., 
corresponding to an increase of the lattice volume by 11 \%. This 
transition persists even if the values of $U$ and $J$ are changed, 
as seen in Fig.~\ref{Fig_2}. As expected, a stronger Coulomb 
repulsion $U$ between the electrons leads to an increase of the 
equilibrium lattice volume. We also observe that for larger $U$ 
values the phase transition occurs at lower volumes. In any case, 
the ncsc and csc calculations both predict a lattice and magnetic 
anomaly upon expansion of the unit cell volume. This anomaly is not 
found in spin polarized DFT calculations for the $(\pi,0)$ and 
$(\pi,\pi)$ antiferromagnetic configurations of iron moments \cite{FeSe_magnetic_DFT}, 
demonstrating the importance of electronic correlations in FeSe.

\subsection{Spectral properties}

\begin{figure}[t]
\centering 
\includegraphics[width=0.35\textwidth,clip=true,angle=-90]{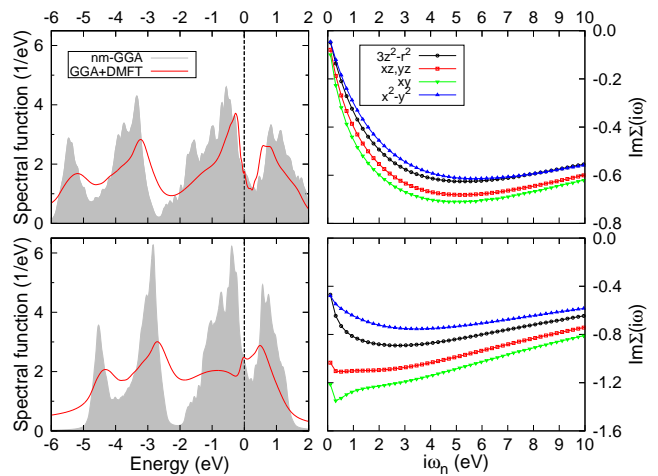}
\caption{(Color online)
Left panels: Spectral functions of paramagnetic FeSe calculated 
within the charge self-consistent DFT+DMFT method (lines) in 
comparison with the DFT results (filled areas). Right panels: 
imaginary parts of the orbitals contributing to the self-energies 
on the Matsubara grid. Top row shows the results obtained for 
$a=7.05$ a.u. Bottom row corresponds to $a=7.6$ a.u.
}
\label{Fig_3}
\end{figure}

To explore the mechanism behind this unusual volume dependence we 
calculate the spectral properties of FeSe and compare the results 
with those obtained from the nsc DFT+DMFT calculations reported 
earlier~\cite{FeSe_Leonov_2015}. The spectral functions computed
at the equilibrium volume ($a=7.05$~a.u.) and above the transition 
($a=7.6$~a.u.) are shown in Fig.~\ref{Fig_3}. Our results overall 
agree with those presented in Ref.~\onlinecite{FeSe_Leonov_2015}. 
In particular, we find a substantial renormalization of the Fe $3d$ 
bands with respect to the DFT results. Indeed, such a behavior is 
common for the pnictides and chalcogenides and is in agreement with 
previous DFT+DMFT results for FeSe \cite{FeSe_Aichhorn_2010, FeSe_Leonov_2015}.
Upon expansion of the lattice, we observe a strong redistribution 
of the spectral weight. In particular, it is seen that the sharp 
peak at -0.19 eV below the Fermi energy in the equilibrium volume 
phase is absent for larger volumes. This peak originates from the 
van Hove singularity of the Fe $xz$/$yz$ and $xy$ bands at the M-point. 
Moreover, for both phases the spectrum shows a broad feature at 
about -1.2 eV which is associated with the lower Hubbard band \cite{FeSe_Aichhorn_2010,WB16}.
The overall change of the spectral function shape upon expansion 
of the lattice agrees well with the evolution of photoemission 
spectra of Fe(Se,Te) series obtained upon increase of the Te 
content \cite{Yokoya}.

\begin{figure}[t]
\centering 
\includegraphics[width=0.47\textwidth,clip=true]{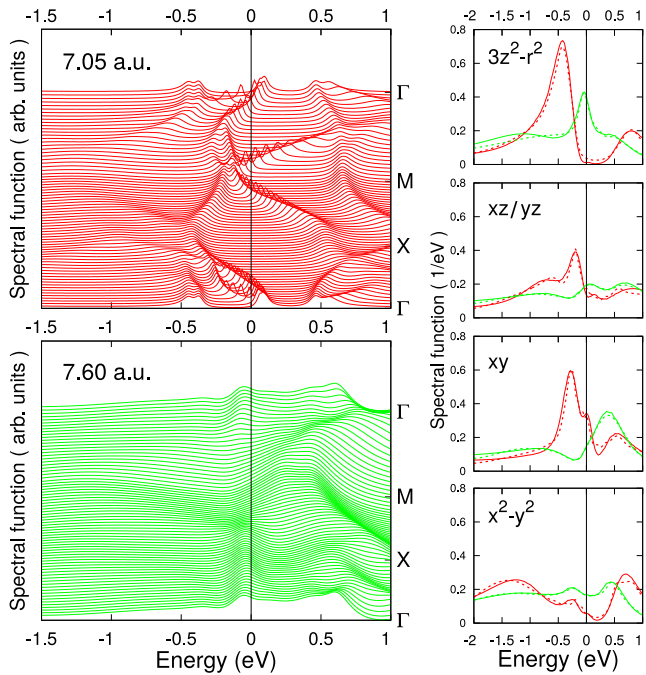}
\caption{(Color online) {\bf k}-resolved spectral functions 
(left column) and partial density of states (right column) 
of paramagnetic FeSe calculated by DFT+DMFT for $a=7.05$ a.u. 
(red) and $7.6$ a.u. (green). The orbitally-resolved contributions 
were evaluated using the maximum entropy method (solid line) 
and Pad\'e approximants (broken line).}
\label{Fig_4}
\end{figure}

\begin{figure}[t]
\centering
\includegraphics[width=0.4\textwidth,clip=true,angle=-90]{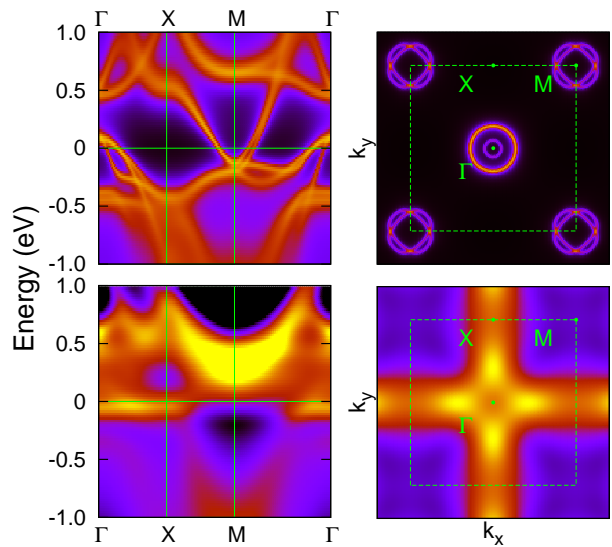}
\caption{(Color online)
Electronic structure (left column) and Fermi surface (right column) of
paramagnetic FeSe in the $\Gamma$-X-M plane of the reciprocal space as
obtained using the charge self-consistent DFT+DMFT method at $T=390$~K.
Top row shows the results for $a=7.05$~a.u. (low volume). Bottom row 
corresponds to $a=7.6$ a.u. (high volume)}
\label{Fig_5}
\end{figure}

Next we calculate the {\bf k}-resolved spectral functions of paramagnetic 
FeSe along the high-symmetry directions of the Brillouin zone. 
In Fig.~\ref{Fig_4} (left panel) we present our results of the 
DFT+DMFT calculations for $a=7.05$ a.u. and $a=7.6$ a.u., respectively. 
The orbitally-resolved integrated spectral functions are shown 
in the right panel of Fig.~\ref{Fig_4}. Our results for the electronic 
structure of FeSe are summarized in the left column of Fig.~\ref{Fig_5}.
Upon expansion of the lattice, we observe a remarkable reconstruction 
of the electronic structure of FeSe (see Figs.~\ref{Fig_4} and 
\ref{Fig_5}) which cannot be described by a simple rescaling or a shift 
of the non-correlated DFT band structure. We find that a substantial
part of the spectral weight in the vicinity of $E_{\mathrm F}$ at 
the M-point is pushed from below to above the Fermi level, while 
the position of the energy bands near the $\Gamma$-point remains
unaffected. This is associated with a correlation-induced shift of 
the van Hove singularity at the M-point above the Fermi level and 
implies an enhancement of the effect of electron correlations upon 
expansion of the lattice of FeSe. We also note that the correlation 
effects exhibit a pronounced orbital-dependent character.

To analyze this behavior in more detail we evaluate the Fermi surface 
of paramagnetic FeSe. In Fig.~\ref{Fig_5} (right column) we display 
the contour map of the spectral weight for the plane $k_z=0$ obtained 
by integration of the spectral function $A({\bf k},\omega)$ over a 
small energy window (5 meV) around the Fermi level. Our results for 
the low-volume phase indicate a well-defined (coherent) Fermi surface 
(FS) which is similar to that in the pnictides \cite{Nesting_fepn}.
The FS exhibits two elliptic electron-like pockets at the M-point and
two circular concentric hole pockets at the $\Gamma$-point. Similar to 
the results obtained from the ncsc calculations \cite{FeSe_Leonov_2015}
the computed FS is characterized by a $(\pi,\pi)$ nesting vector
connecting the electron and hole sheets. A comparison of the calculated 
FS of paramagnetic FeSe with experiment shows that the size of the 
measured FS pocket is smaller than that obtained within DFT+DMFT. 
This is in accordance with previous DFT and DFT+DMFT studies \cite{WB16,WK16},
suggesting, e.g., the importance of non-local correlations effects, 
frustration magnetism \cite{GM15}, or spin-orbit interaction effects \cite{BE16}.
Upon expansion of the lattice, we observe an abrupt change of the
topology of the Fermi surface (Lifshitz transition). In particular, 
the spectral weight of the electron pockets centered at the M-point 
vanishes. The hole pocket encircling the $\Gamma$-point transforms 
into a large square-like FS surrounding the M-point with the four 
pronounced spots around the $\Gamma$-point. This transition results 
in a change of the dominating nesting vector from $(\pi,\pi)$ to 
$(\pi,0)$. The observed topological change proceeds similar to the 
evolution of the experimental photoemission spectra \cite{FeSe_photoemission_2, FeSe_photoemission_3} 
of the doped FeSe$_{1-x}$Te$_{x}$ samples. These data confirm the 
emergence of the Fermi surface pocket at the X-point for large 
concentrations of Te for $x>0.7$.

\subsection{Orbital-selective renormalization}
\begin{figure}[t]
\centering 
\includegraphics[width=0.4\textwidth,clip=true]{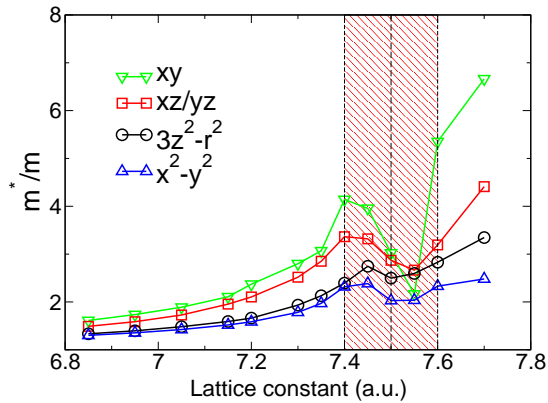}
\caption{(Color online) Orbitally-resolved quasiparticle mass 
enhancement $m^{*}/m$ of the Fe $3d$ states in paramagnetic 
FeSe as a function of lattice constant calculated by the charge 
self-consistent DFT+DMFT approach. The critical region associated 
with the electronic and structural transition is indicated by a 
red filled rectangle.
}
\label{Fig_6}
\end{figure}

An expansion of the lattice also goes along with a remarkable 
orbital-selective renormalization of the Fe $3d$ bands (see Fig~\ref{Fig_6}),
indicating significantly stronger renormalization of the $t_{2}$ 
bands ($xz/yz$ and $xy$) than of the $e$ bands ($3z^2-r^2$ and $x^2-y^2$). 
In Fig.~\ref{Fig_3} (right column) we show the Fe $3d$ imaginary
self-energies for the low- and high-volume phases, respectively. 
At the equilibrium volume, the self-energy obeys a Fermi-liquid-like 
behavior characterized by a weak damping of quasiparticles. By contrast, 
the expanded-volume phase shows a pronounced orbital-selective
behavior, associated with a non-Fermi-liquid behavior of the $t_{2}$
orbitals. Indeed, the self-energies of the $t_{2}$ orbitals decrease 
with decreasing Matsubara frequency -- and in the case of the 
self-energy of the $xy$-orbital even seems to diverge -- but finally 
show an upturn at the lowest Matsubara frequency. At the same time, 
the $e$ states remain Fermi-liquid-like, but with a damping which 
is about five times stronger than that in the equilibrium phase. 
These results agree well with an analysis of the band mass enhancement
${m^*}/{m}=1-\partial \mathrm{ Im }\Sigma(\omega)/\partial\omega|_{\omega=0}$,
which provides a quantitative measure of the correlation strength. 
In Fig.~\ref{Fig_6} we display the computed mass enhancement $m^*/m$ 
as a function of lattice constant. In the vicinity of the equilibrium 
lattice constant $m^{*}/m$ lies in the range $1.5$--$2$. Upon 
expansion of the lattice it shows a substantial increase followed 
by a critical region at $a\sim 7.5$ a.u. (where the electronic and 
structural transition occurs), which is characterized  by a change 
of the sign of its derivative. Furthermore, the effective mass of 
the $t_{2}$ electrons exhibit larger renormalizations than in the
$e$ orbitals. Indeed, for the former it reaches $\sim$ 6.5 and 4.5 
for the Fe $xy$ and $xz/yz$ states, respectively.

\begin{figure}[t]
\centering
\includegraphics[width=0.47\textwidth,clip=true]{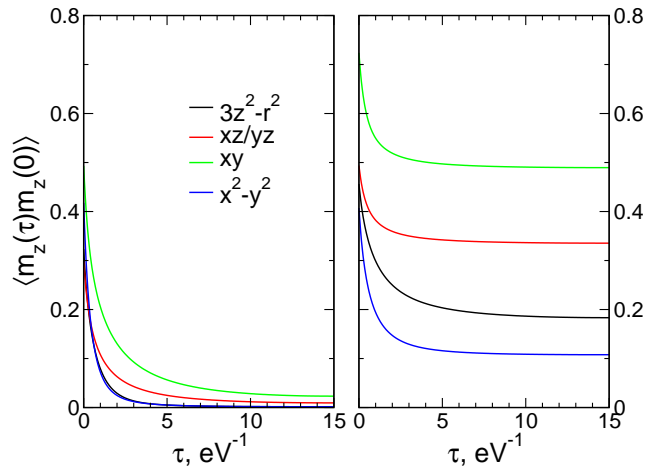}
\caption{(Color online) Orbitally-resolved local spin correlation 
functions $\chi(\tau) = \langle \hat{m}_z(\tau) \hat{m}_z(0)\rangle$ 
of paramagnetic FeSe calculated using DFT+DMFT for the lattice constant 
$a=7.05$ a.u. (left) and $a=7.60$ a.u. (right).
}
\label{Fig_7}
\end{figure}

%%%%%%%%%%%%%%%%%%%%%%%%%%%%%%%%%%%%%%%%%%%%%%%%%%%%%%%%%%%%%%%%%%%%%%%%%%

\subsection{Susceptibility}
The electronic and structural phase transition is accompanied by a 
significant growth of the fluctuating local magnetic moment (see 
lower panel of Fig.~\ref{Fig_1}). The transition is found to result 
in a crossover from an itinerant to localized moment behavior, as it 
is seen from the local spin susceptibility
$\chi(\tau ) = \langle \hat{m}_z(\tau) \hat{m}_z(0) \rangle$, where 
$\tau$ is the imaginary time. The results for the different orbital 
contributions are presented in Fig. \ref{Fig_7}. This behavior is 
consistent with the coherence-incoherence transition scenario which 
was found experimentally in the Fe(Se,Te) series~\cite{FeSe_photoemission_3}. 
Moreover, our calculations reveal a strong orbital-selectivity in 
the formation of the local moments upon expansion of the lattice of 
FeSe. Here the $xy$ orbital plays a predominant role, while the 
contribution of the $xz/yz$ orbitals is substantially weaker. On the 
other hand, the $e$ orbitals exhibit an itinerant moment behavior.

\begin{figure}[t]
\centering
\includegraphics[width=0.2\textwidth,clip=true,angle=-90]{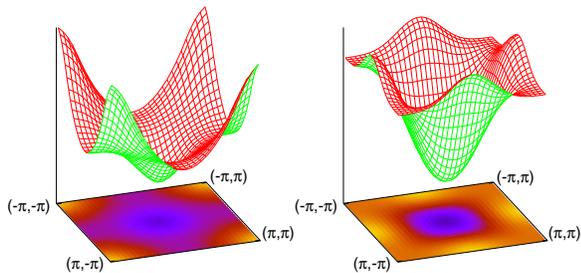}
\caption{(Color online)
Momentum dependence of the local spin susceptibility $\chi({\bf q})$ 
of paramagnetic FeSe calculated using the charge self-consistent 
DFT+DMFT method for $a=7.05$ a.u. (left) and $a=7.6$ a.u. (right).}
\label{Fig_8}
\end{figure}

In addition, we compute the momentum-dependent local spin 
susceptibility $\chi({\bf q})$ in the $(q_x,q_y)$ plane for $q_z=0$. 
Our results are presented in Fig.~\ref{Fig_8}. The susceptibility 
calculated for the equilibrium volume shows a maximum at the 
corners of the tetragonal Brilloiun zone at the M-points. This 
confirms that the leading magnetic instability of FeSe at ambient 
pressure occurs at the wave vector $(\pi,\pi)$, in agreement with 
experiment \cite{FeSe_pi_pi_fluctuations}. An expansion of the 
lattice volume leads to a dramatic change of $\chi({\bf q})$, 
associated with a suppression of the maximum at $(\pi,\pi)$ and 
the development of a maximum at $(\pi,0)$. This change of the 
magnetic correlations is associated with the change of the Fermi 
surface (Lifshitz transition) discussed above. The evolution of 
$\chi({\bf q})$ qualitatively agrees with the experimentally 
observed transformation of magnetic correlations in the Fe(Se,Te) 
series \cite{FeSe_magnetism}. Indeed, our results show a transition 
from $(\pi,\pi)$-type antiferromagnetic fluctuations in the 
paramagnetic tetragonal phase of FeSe to $(\pi,0)$-type magnetism 
upon expansion of the lattice.

\begin{figure}[t]
\centering
\includegraphics[width=0.47\textwidth,clip=true]{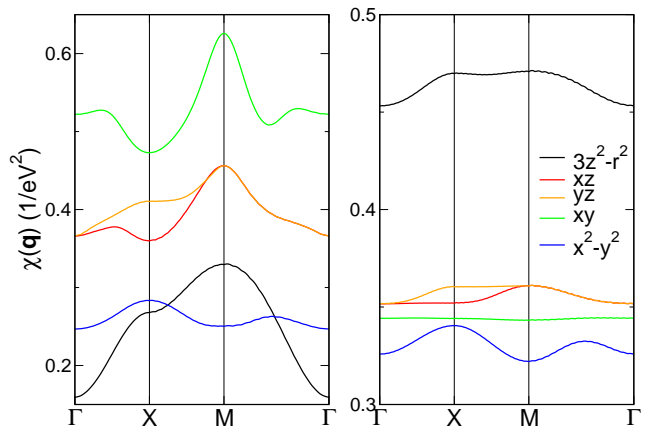}
\caption{(Color online) Orbitally-resolved local spin susceptibility 
$\chi({\bf q})$ of paramagnetic FeSe calculated along the $\Gamma$-X-M-$\Gamma$ 
path using the charge self-consistent DFT+DMFT for $a=7.05$ a.u. (left) 
and $a=7.6$ a.u. (right).}
\label{Fig_9}
\end{figure}

Moreover, we calculate the orbital contributions of $\chi({\bf q})$
along the $\Gamma$-X-M-$\Gamma$ path (Fig.~\ref{Fig_9}). For $a=7.05$~a.u. 
we observe a strong orbital-selective behavior of magnetic correlations 
with a leading contribution originating from the Fe $xy$ orbital. This 
orbital leads to a maximum of $\chi({\bf q})$ at the M-point, confirming 
that magnetic correlations in FeSe are predominantly of the $(\pi,\pi)$-type.
On the other hand, the behavior of $\chi({\bf q})$ in the high volume
phase is completely different. In particular, for $a=7.6$~a.u. the leading
contribution to $\chi({\bf q})$ is due to the Fe $3z^2-r^2$ orbital which 
varies only weakly along the $\Gamma$-X-M-$\Gamma$ path. Our analysis shows 
that only the inclusion of all orbital contributions (especially of the 
$x^2-y^2$ orbital contribution, which exhibits the most substantial variation 
in the reciprocal space and shows a maximum at the X-point) results in
the $(\pi,0)$-type magnetic correlations prevalent in the high-volume
phase of Fe(Se,Te).

Our results for $\chi({\bf q})$ in Fig.~\ref{Fig_9} demonstrate that 
for $a=7.05$ a.u. the $xz$ and $yz$ orbitals contribute very differently 
to $\chi( {\bf q} )$ along the $\Gamma$-X-M direction. It will be 
interesting to check whether this finding, together with the symmetry-induced 
splitting between the $xz/yz$ orbitals at the X point, can stabilize 
the observed nematicity in FeSe \cite{nematicity}, for example through 
the coupling of magnetic fluctuations to phonons near the X point.

%%%%%%%%%%%%%%%%%%%%%%%%%%%%%%%%%%%%%%%%%%%%%%%%%%%%%%%%%%%%%%%%%%%%%%%%%%%%%%%%%%%%%%%%%%%%%%%%
%%%%%%%%%%%%%%%%%%%%%%%%%%%%%%%%%%%%%%%%%%%%%%%%%%%%%%%%%%%%%%%%%%%%%%%%%%%%%%%%%%%%%%%%%%%%%%%%

\section{Conclusion}
In conclusion, we studied the electronic structure and phase 
stability of the tetragonal paramagnetic phase of FeSe using 
a fully charge self-consistent implementation of the DFT+DMFT 
method. Our results demonstrate the importance of electron 
correlation effects which, in particular, trigger the anomalous 
behavior of FeSe upon expansion of the lattice volume. We note 
that such an expansion can be experimentally realized by the 
isovalent substitution of Se with Te. Our results also reveal 
a complete change of the electronic structure of paramagnetic 
FeSe upon a moderate expansion of the lattice (at -7.6 GPa). 
This behavior is associated with a remarkable reconstruction 
of the Fermi surface topology (Lifshitz transition) of FeSe and 
is accompanied by a change of the in-plane magnetic nesting 
vector from $(\pi,\pi)$ to $(\pi,0)$, in agreement with experiment \cite{FeSe_magnetism}.
This behavior is intimately linked with an orbital-selective 
transition from itinerant to localized moment behavior, where 
the Fe $xy$ orbitals contribute most strongly. The phase transformation 
is driven by a correlation-induced shift of the van Hove 
singularity of the Fe $t_{2}$ bands at the M-point across the 
Fermi level \cite{CE15}. We also observe a strong orbital-selective 
renormalization of the Fe $3d$ band structure, with the largest 
contribution coming again from the Fe $xy$ orbital, which gives 
rise to a non-Fermi-liquid-like behavior above the transition. \cite{WS14}
In view of our results the complex behavior of the chalcogenide 
parent system Fe(Se,Te), such as the anomalous increase of the 
superconducting temperature upon positive or negative pressure, 
appears to be associated with the proximity of the van Hove 
singularity of the Fe $t_{2}$ bands at the M-point to the Fermi 
level, and with the sensitivity of its position to external 
conditions \cite{CE15}. Furthermore, our results for the local 
spin susceptibility $\chi({\bf q})$, which exhibits a strong 
splitting between the $xz$ and $yz$ orbitals near the X-point, 
suggest a spin-fluctuation origin of the nematic phase of 
paramagnetic FeSe. This will be the subject of further investigations.

%%%%%%%%%%%%%%%%%%%%%%%%%%%%%%%%%%%%%%%%%%%%%%%%%%%%%%%%%%%%%%%%%%%%%%%%%%%%%%%%%%%%%%%%%%%%%%%%

\section{Acknowledgments}
We thank V. Tsurkan, J. Schmalian, and L. H. Tjeng for useful discussions.
I.L. acknowledges support from the Deutsche Forschungsgemeinschaft
through Transregio TRR 80 and the Ministry of Education and Science
of the Russian Federation in the framework of Increase Competitiveness
Program of NUST "MISIS" (K3-2016-027), implemented by a governmental
decree dated 16th of March 2013, N 211. D.V., S.L.S. and V.I.A. are
grateful to the Deutsche Forschungsgemeinschaft for financial support
through the Research Unit FOR 1346. 

%%%%%%%%%%%%%%%%%%%%%%%%%%%%%%%%%%%%%%%%%%%%%%%%%%%%%%%%%%%%%%%%%%%%%%%%%%%%%%%%%%%%%%%%%%%%%%%%
%%%%%%%%%%%%%%%%%%%%%%%%%%%%%%%%%%%%%%%%%%%%%%%%%%%%%%%%%%%%%%%%%%%%%%%%%%%%%%%%%%%%%%%%%%%%%%%%


\begin{thebibliography}{99}
\bibitem{note}
The fluctuating local magnetic moment is evaluated from the imaginary-time average of the
local spin susceptibility $\chi(\tau) = \langle \hat{m}_z(\tau) \hat{m}_z(0)\rangle$ as
$M_z = k_BT\int_0^{1/k_BT}\chi(\tau)~d\tau$, where $\tau$ is the imaginary time.
\end{thebibliography}
\end{document}